\documentclass[preprint,11pt,aps,prd,nofootinbib]{revtex4-1}
\usepackage{amsmath}
\usepackage{amssymb}
\usepackage{color}
\usepackage{hyperref}
\usepackage{mathtools}

\begin{document}

\title{Deformed relativity symmetries and the local structure of spacetime}
\author{Marco Letizia}
\email{mletizia@sissa.it}
\author{Stefano Liberati}
\email{liberati@sissa.it}
\affiliation{SISSA, Via Bonomea 265, 34136 Trieste, Italy and INFN, Sez. di Trieste}

\begin{abstract}

A spacetime interpretation of deformed relativity symmetry groups was recently proposed by resorting to Finslerian geometries, seen as the outcome of a continuous limit endowed with first order corrections from the quantum gravity regime. In this work we further investigate such connection between deformed algebras and Finslerian geometries by showing that the Finsler geometries associated to the generalisation of the Poincar\'{e} group (the so called $\kappa$-Poincar\'{e} Hopf algebra) are maximally symmetric spacetimes which are also of the Berwald type: Finslerian spacetimes for which the connections are substantially Riemannian, belonging to the unique class for which the weak equivalence principle still holds. We also extend this analysis by considering a generalization of the de Sitter group (the so called $q$-de Sitter group) and showing that its associated Finslerian geometry reproduces locally the one from the $\kappa$-Poincar\'{e} group and that itself can be recast in a Berwald form in an appropriate limit.
\end{abstract}

\maketitle
\tableofcontents

\section{Introduction}

The Weak Equivalence Principle (WEP) and local Lorentz invariance (LI) represent the building blocks of metric theories of gravity being at the basis of the Einstein Equivalence Principle (together with the local position invariance) \cite{lrr-2006-3}.

Departure from standard LI have been nowadays considered in essentially all Quantum Gravity (QG) scenarios as it represents a major source of phenomenological investigations (see \cite{Liberati:2013xla}). At least three ways of dealing with LI at high energy have been taken into account in the literature: preservation of standard Lorentz invariance at every energy scale (for instance, in Causal Set Theory \cite{Dowker:aza}), hard Lorentz invariance violations (LIVs) at high energy (e.g., Ho\v{r}ava--Lifshitz gravity \cite{Horava:2009uw}) and deformations of the relativity group (or deformed special relativity, DSR \cite{AmelinoCamelia:2000mn}), where the relativity principle is preserved and a new invariant mass scale (often taken to be the Planck mass) is introduced beside the speed of light.

In all of these possibilities a crucial role is played by modified dispersion relations (MDR) for elementary particles, that can generically be written in the following form
\begin{equation}
E^2=m^2+p^2+\sum_{n=1}^\infty a_n (\mu,M) p^n,
\end{equation}
where $p=\sqrt{|\vec{p}|^2}$, $a_n$ are dimensional coefficients, $\mu$ is some particle physics mass scale and $M$ is the mass scale characterizing the physics responsible for the departure from standard LI (usually identified with the Planck mass).

While the most stringent constraints have been put on coefficients associated with Lorentz violating operators (in an effective field theory approach), the same cannot be said about effects related to DSR which, furthermore, have been mostly described within momentum space in the Hamiltonian formalism.

In the context of gravitational theories, a way to introduce LIV is to consider models in which the ``ground state" configuration possesses less symmetries than Minkowski spacetime. This is the case, for example, of Einstein--Aether theory (see \cite{Jacobson:2008aj}), where such a state is given by the Minkowski spacetime with a fixed norm timelike vector field that breaks boost invariance. On the other hand, in an attempt to provide a spacetime description of the deformed symmetries a la DSR, it would be interesting to understand if these can be related to some new local structure of spacetime, possibly described by some maximally symmetric background generalising a pseudo-Riemannian structure and Minkowski spacetime.

A concrete example of (quantum) deformation of the ordinary Poincar\'{e} group is represented by the $\kappa$-Poincar\'{e} ($\kappa$P) Hopf algebra
\cite{Majid:1994cy,LUKIERSKI1991331,Lukierski:1993wxa}. $\kappa$P symmetries have been shown to characterize the kinematics of particles living on a flat spacetime and non-trivial momentum space with a de Sitter geometry \cite{Gubitosi:2013rna,AmelinoCamelia:2011nt,KowalskiGlikman:2003we,KowalskiGlikman:2002ft} and they have been shown to naturally emerge in the context of $2+1$ dimensional QG coupled to point particles (see e.g. \cite{Freidel:2003sp}).

In some recent papers \cite{Girelli:2006fw,Amelino-Camelia:2014rga} a strong link was established between the momentum space analysis, usually carried out when dealing with $\kappa$P symmetries (and the associated MDR), and the spacetime picture provided by Finsler geometry (a generalisation of Riemannian geometry whose properties will be reviewed in Sec.\ref{finslersec}).\footnote{See also \cite{Lobo:2016lxm} for a spacetime description of particles with MDR and the relation between the spacetime metric and the momentum space metric.} The Finsler geometry associated with $\kappa$P represents an instance of the kind of spacetimes we were discussing earlier, i.e., a flat maximally symmetric spacetime that is not Minkowski. 

Among all the possible Finsler structures, a particular case is given by Berwald spaces. These are the Finsler spaces that are the closest to be Riemmanian (we will provide a more precise definition in Sec.\eqref{finslersec}). If a Finsler space is of the Berwald type then any observer in free fall looking at neighbouring test particles would observe them move uniformly over straight lines accordingly to the Weak Equivalence Principle~\cite{Minguzzi:2016gct}. Interestingly enough the Finsler metric correspondent to $\kappa$P symmetries found in~\cite{Amelino-Camelia:2014rga} appears to be a member of this class. However, we shall see that this come about in a somewhat trivial way as a straightforward consequence of the flatness of the metric in coordinate space.
With this in mind, it would be interesting to consider examples of curved metrics associated to more general deformed algebras so to check if for these the local structure of spacetime does not reduce to the Minkowski spacetime but rather to the Finsler geometry with $\kappa$P symmetries and furthermore for checking if also these geometries are of the Berwald type. 

Missing a definitive derivation of such hypothetical curved-deformed-geometries based on some quantum gravity model, one has to resort also in this case to study a case for which a deformed group of symmetry is available and a Finslerian metric can be derived. In this sense a case of particular interest is the $q$-de Sitter ($q$dS) Hopf algebra \cite{0305-4470-26-21-019,Lukierski:1991ff}, a quantum deformation of the algebra of isometries of the de Sitter spacetime. It represents a case in which curvature of momentum space is present together with curvature in spacetime in the context of a well defined relativistic framework.\footnote{See \cite{PhysRevD.92.084053} for a description of particles with modified dispersion relation in the context of Hamilton geometry.} As such, this represents the perfect arena for our analysis.\footnote{During the final stage in the preparation of this paper, the work \cite{Lobo:2016xzq} has been released on the arXiv...}

Let us stress that such models of Finslerain spacetimes, embodying new group of symmetries, do not have to be considered as definitive proposals for the description of quantum gravitational phenomena at a fundamental level. We take here the point of view for which, between the full quantum gravity regime and the classical one, there is an intermediate phase where a continuous spacetime can be described in a \textit{semi-classical} fashion. In particular, if the underlying QG theory predicts that spacetime is in some way discrete, then we assume that a meaningful continuum limit can be performed and that this limit is not equivalent to a classical limit. The outcome of this hypothetical procedure would be a spacetime that can be described as \textit{continuum} but still retaining some quantum features of the fundamental theory. Then the departure from the purely classical theory will be weighed by a \textit{non-classicality} parameter (potentially involving the scale of Lorentz breaking/deformation) and in the limit in which this parameter goes to zero, the completely classical description of spacetime is recovered (see e.g.~\cite{Assaniousssi:2014ota,Torrome:2015cga} for a concrete example of such a construction).

The purpose of this paper is then twofold. In the first part we will show that there exists a Finsler spacetime associated to the mass Casimir of $q$dS following a procedure that is analogous to the one presented in \cite{Girelli:2006fw} and explicitly compute the associate Finsler metric and Christoffel symbols. We will then discuss how, in the limit in which the curvature goes to zero, one recovers the Finsler structure of $\kappa$P thus providing an example of a curved Finsler spacetime whose local limit is not trivially given by the Minkowski spacetime. In the second part we will discuss how, in a particular limit, the Finsler structure associated with $q$dS becomes of the Berwald type. Finally, we will discuss what are the consequences of these results and speculate about possible phenomenological studies.

\section{Finsler geometry and modified dispersion relations}\label{finslersec}

Let us begin by reviewing some basics notions concerning Finsler geometry, loosely following \cite{Girelli:2006fw}. Given a manifold $M$ of dimension $D$, Finsler geometry is a generalization of Riemannian geometry where, instead of defining an inner product structure over the tangent bundle $TM$, one defines a norm $F$. This norm is a real function $F(x,v)$, with $v\in T_x M$ (the tangent space at the point $x$ of the manifold), and it satisfies the following properties:
\begin{itemize}
\item $F(x,v)\neq 0$\quad for $v\neq 0$,
\item $F(x,\alpha v)=\left|\alpha\right| F(x,v)$\quad for $\alpha\in\mathbb{R}$.
\end{itemize}
The Finsler metric can be defined as
\begin{equation}\label{fmetrdef}
g_{\mu\nu}(x,v)=\frac{1}{2}\frac{\partial^2 F^2(x,v)}{\partial v^\mu\partial v^\nu},
\end{equation}
and, using the Euler theorem for homogeneous functions, it can be shown that the relation above is equivalent to
\begin{equation}
F(x,v)=\sqrt{g_{\mu\nu}(x,v)v^\mu v^\nu}.
\end{equation}
Therefore $g_{\mu\nu}(x,v)$ is homogeneous of degree zero in $v$. Given that, by definition, $g_{\mu\nu}$ is non degenerate, the inverse exists and it satisfies $g_{\mu\nu}(x,v)g^{\nu\rho}(x,v)=\delta_\mu^\rho$. Moreover, since $F^2$ is a homogeneous function of degree two in the velocities, the metric satisfies the following relations
\begin{equation}\label{fprop}
v^\alpha\frac{\partial g_{\mu\nu}}{\partial v^\alpha}=v^\mu\frac{\partial g_{\mu\nu}}{\partial v^\alpha}=v^\nu\frac{\partial g_{\mu\nu}}{\partial v^\alpha}=0,
\end{equation}It is clear that the Riemannian case is obtained when $F$ is quadratic in $v$ and it is defined by an inner product with a velocity independent metric tensor. Using the norm $F$, one can naturally introduce a notion of distance. Indeed, as in Riemannian geometry, one can define the length of an arc of curve as
\begin{equation}\label{arclength}
\ell (\mathcal{C})=\int_{\tau_1}^{\tau_2} F(x,v)d\tau,
\end{equation}
where $\tau$ is a parameter for the curve $\mathcal{C}$. Because of the homogeneity properties of the norm, the expression above is reparametrization invariant. Taking two points $p,q\in M$, and considering all the curves connecting these two points, the minimum of the lengths of all these curves defines a distance between the two points.

This construction is usually carried out in Euclidean signature. We will assume that the extension to Lorentzian signature (Finsler spacetimes or pseudo-Finsler geometry) can be done, as that is the case in many examples (see anyway the appendix of \cite{Weinfurtner:2006wt}). From now on, for the sake of brevity, we will omit the prefix ``pseudo" and we will always implicitly consider the Lorentzian case.

\subsection{Derivation of Finsler geometries from modified dispersion relations}

In this section we review the procedure introduced in \cite{Girelli:2006fw} for deriving the Finsler geometry associated with a particle with a modified dispersion relation.

Let us start by considering the action of a particle with a constraint imposing the on-shell relation $\mathcal{M}(p)=m^2$
\begin{equation}
I=\int \left[\dot{x}^\mu p_\mu-\lambda\left(\mathcal{M}(p)-m^2\right)\right]d\tau,
\end{equation}
where $\lambda$ is a Lagrange multiplier. In order to find the explicit expression of the Lagrangian we use Hamilton's equations that read as
\begin{equation}
p_\mu=\lambda\frac{\partial\mathcal{M}}{\dot{x}^\mu}.
\end{equation}
If the relation above is invertible, one is able to rewrite the action in terms of velocities and the multiplier hence obtaining\footnote{The symbols $x$ and $\dot{x}$, when taken as arguments of functions, generically refer to both the time and spatial component of the coordinates and the velocities.}
\begin{equation}
I=\int \mathcal{L}(x,\dot{x},\lambda)d\tau.
\end{equation}
We can also eliminate the multiplier using the equation of motion obtained varying the action with respect to it so to get the expression of the Lagrangian in terms of velocities only  $\mathcal{L}(x,\dot{x},\lambda(x,\dot{x}))$.

Finally we can identify the Finsler norm through the following relation
\begin{equation}\label{fnorm}
\mathcal{L}(x,\dot{x},\lambda(x,\dot{x}))=mF(x,\dot{x}),
\end{equation}
and the Finsler metric is then given by the Hessian matrix of $F^2$ as in \eqref{fmetrdef}.
At this point, the action can be written as
\begin{equation}\label{finsleraction}
I=m\int F d\tau=m\int\sqrt{g_{\mu\nu}(x,\dot{x})\dot{x}^\mu\dot{x}^\nu}d\tau,
\end{equation}
which correspond to the action of a free relativistic particle propagating on a spacetime described by a velocity dependent metric.

Using the definition of generalized momentum, one can now simply relate the four-momentum to the Finsler norm as
\begin{equation}
p_\mu=m\frac{\partial F}{\partial \dot{x}^\mu}=m\frac{g_{\mu\nu}\dot{x}^\nu}{F}.
\end{equation}
Moreover using the inverse metric $g^{\mu\nu}$ one recovers the dispersion relation in a simple way as
\begin{equation}
m^2=g^{\mu\nu}(\dot{x}(p))p_\mu p_\nu.
\end{equation}

In the last parts of this section, we will introduce the notion of Berwald spaces and we will recap some known results regarding the Finsler structure associated with the $\kappa$P group.

\subsection{Berwald spaces and normal coordinates}\label{berwald}

\textit{Berwald spaces are Finsler spaces that are just a bit more general than Riemannian and locally Minkowskian spaces. They provide examples that are more properly Finslerian, but only slightly so} \cite{bao2000introduction}.

The statement above is a good intuitive description of what Berwald spaces are. One of the (equivalent) technical characterizations of Berwald spaces is the following \cite{bao2000introduction}:

\begin{itemize}
\item The quantities $\partial^2_{\dot{x}}(G^\mu)$, with $G^\mu:=\Gamma^\mu_{\,\rho\sigma}(x,\dot{x})\dot{x}^\rho \dot{x}^\sigma$, do not depend on $\dot{x}^\mu$.
\end{itemize}

The objects $\Gamma^\mu_{\,\rho\sigma}$ are the usual Christoffel symbols defined as
\begin{equation}
\Gamma^\mu_{\rho\sigma}(x,\dot{x})=\frac{1}{2}g^{\mu\nu}\left(\partial_\rho g_{\sigma\nu}+\partial_\sigma g_{\rho\nu}-\partial_\nu g_{\rho\sigma}\right),
\end{equation}
that for a general Finsler metric depend on $\dot{x}^\mu$. The coefficients $G^\mu$ are called spray coefficients and they appear in the geodesic equations, obtained by minimizing the action \eqref{arclength} (or \eqref{finsleraction}), as
\begin{equation}
\ddot{x}^\sigma+2\,G^\sigma=\frac{\dot{F}}{F}\dot{x}^\sigma,
\end{equation}
where the right-hand side is vanishing for a constant speed parametrization. In other words a Finsler space is of the Berwald type when the $G^\mu$ are purely quadratic in the velocities\footnote{For an introduction to the various kind of connections that one can define in Finsler geometry see \citep{bao2000introduction}.}.


In pseudo-Riemannian geometry, normal coordinates can be defined in a neighbourhood of a point $p$ (Fermi coordinates along a curve $\gamma$), such that the Christoffel symbols of the connection vanish at $p$ (along $\gamma$) \cite{Manasse1963}. This procedure fails in pseudo-Finsler geometry if the space is not Berwald (see \cite{Busemann1955}, \citep{bao2000introduction} and references therein). Therefore in Finsler geometry the existence of free falling observers looking at nearby free falling particles moving in straight lines is not guaranteed. In this respect, Berwald spaces play an important role in determining whether a given Finsler structure violates the Weak Equivalence Principle (WEP).

\subsection{Results for $\kappa$-Poincar\'{e}}

In this subsection we will briefly review the results obtained in \citep{Amelino-Camelia:2014rga}, regarding the Finsler structure associated with the $\kappa$P group. The mass Casimir of the $\kappa$P algebra, at first order in the deformation parameter $\ell$, is given by\footnote{The $\kappa$P algebra can be derived from the $q$dS algebra in an appropriate limit (see Sec.~\ref{qdS}). See also \cite{Gubitosi:2013rna}.}
\begin{equation}
C_\ell=p_0^2-p_1^2\left(1+\ell p_0\right).
\end{equation}
Following the procedure outlined in the previous subsection, the associated Finsler norm reads as
\begin{equation}
F_\ell=\sqrt{\dot{t}^2-\dot{x}^2}+\frac{\ell m}{2}\frac{\dot{t}\dot{x}^2}{\dot{t}^2-\dot{x}^2},
\end{equation}
and, using \eqref{fmetrdef}, the Finsler metric is
\begin{equation}\label{kmetric}
g^{F_\ell}_{\mu\nu}(x,\dot{x})=\left(
\begin{array}{cc}
 1+\frac{3 m \ell  \dot{t} \dot{x}^4}{2 \left(\dot{t}^2- \dot{x}^2\right)^{5/2}} 
 & \frac{m \ell \dot{x}^3\left(\dot{x}^2-4 \dot{t}^2 \right)}{2 \left(\dot{t}^2-\dot{x}^2\right)^{5/2}} \\
 \frac{m \ell \dot{x}^3\left(\dot{x}^2-4 \dot{t}^2 \right)}{2 \left(\dot{t}^2-\dot{x}^2\right)^{5/2}} 
 & -1+\frac{m \ell\dot{t}^3 \left(2 \dot{t}^2+\dot{x}^2 \right)}{2 \left(\dot{t}^2-\dot{x}^2\right)^{5/2}} \\
\end{array}
\right).
\end{equation}
It can be easily checked that the metric above satisfies the relations \eqref{fprop} and that it can be rewritten in momentum space as follows
\begin{equation}
g^{F_\ell}_{\mu\nu}(x,p)=\left(
\begin{array}{cc}
 1+\frac{3}{2}\frac{\ell p_0 p_1^4}{m^4} 
 & -\frac{\ell}{2}\frac{p_1^3\left(p_1^2-4p_0^2\right)}{m^4} \\
  -\frac{\ell}{2}\frac{p_1^3\left(p_1^2-4p_0^2\right)}{m^4} 
 & -1+\frac{\ell}{2}\frac{p_0^3\left(2p_0^2+p_1^2\right)}{m^4} \\
\end{array}
\right).
\end{equation}
Using this expression, the dispersion relation can be simply given as
\begin{equation}
g_{F_\ell}^{\,\mu\nu}p_\mu p_\nu=p_0^2-p_1^2\left(1+\ell p_0\right).
\end{equation}

In \citep{Amelino-Camelia:2014rga}, it was also shown that the Killing vectors associated with the metric \eqref{kmetric} are compatible with the $\kappa$P symmetries.

Interestingly enought, it can be easily proven that the Finsler metric associated with $\kappa$P has vanishing Christoffel symbols and that the relation $\Gamma^\mu_{\rho\sigma}=0$ trivially satisfies the conditions for a Berwald space. This was expected since, in \citep{Amelino-Camelia:2014rga}, a deformation of a special-relativistic particle was considered and in that case the metric had no dependence on coordinates, meaning that the spacetime geometry was flat.

The subsequent question is whether that was a coincidence or not. In other words, since all locally Minkowskian spacetimes are Finsler spacetimes of the Berwald type \citep{bao2000introduction}, do Berwald spaces play an important role regarding the local structure of spacetime with DSR-like symmetries or it is just a trivial consequence of local flatness? To answer this question we shall then examine the Finsler geometry of a spacetime related to the $q$dS mass Casimir that reduces to the $\kappa$P Finsler geometry when the curvature goes to zero. 

\section{q-de Sitter inspired Finsler spacetime}

In what follows we shall explicitly investigate the Finsler metric associated to a $q$-de Sitter Hopf algebra and consider its local limit to prove that it reproduces the $\kappa$-Poincar\'{e} Finsler geometry. We shall then also investigate if  $q$-de Sitter Finsler geometry is per-se of the Berwald type.

\subsection{$q$-de Sitter}\label{qdS}
Let us start by denoting the key features of the 1+1D $q$dS Hopf algebra \cite{Barcaroli:2015eqe}. Using the notation of \cite{Barcaroli:2015eqe}, the commutators among the symmetry generators are
\begin{align}
&[P_0,P]=HP,\quad[P_0,N]=P-HN,\notag \\
&[P,N]=\cosh(w/2)\frac{1-e^{\frac{-2wP_0}{H}}}{2w/H}-\frac{1}{H}\sinh(w/2)e^{\frac{-wP_0}{H}}\Theta,
\end{align}
where 
\begin{equation}
\Theta=\left[e^{\frac{wP_0}{H}}(P-HN)^2-H^2 e^{\frac{wP_0}{H}}N^2\right],
\end{equation}
and $P_0,P,N$ refer to the generators of time translation, space translation and boost respectively, $H$ is the Hubble rate and $w$ is the deformation parameter.

For the coproducts, which are used to express the conservation of momentum when dealing with multiple particles, one has
\begin{equation}
\Delta(P_0)=1\otimes P_0+P_0\otimes 1,\quad\Delta(P)=e^{\frac{-wP_0}{H}}\otimes P+P\otimes 1,\quad\Delta(N)=e^{\frac{-wP_0}{H}}\otimes N+N\otimes 1,
\end{equation}
while the antipodes are
\begin{equation}
S(P_0)=-P_0,\quad S(P)=e^{\frac{wP_0}{H}} P_1,\quad S(N)=e^{\frac{wP_0}{H}} N.
\end{equation}
Finally the mass Casimir is
\begin{equation}
C_{q\text{dS}}=H^2\frac{\cosh(w/2)}{w^2/4}\sinh^2\left(\frac{w P_0}{2H}\right)-\frac{\sinh(w/2)}{w/2}\Theta.
\end{equation}
The parameter $w$ is usually assumed to be a dimensionless combination of a fundamental length scale $\ell$ and the dS radius $H^{-1}$. There are various possible choices (see for example \cite{Marciano:2010gq}) and we will focus on the one that gives back the classical dS algebra for $\ell\rightarrow 0$ and the $\kappa$P algebra for $H\rightarrow 0$, i.e., $w=H\ell$.

Upon introducing a representation of the phase space coordinates $x^\mu=\left\{t,x\right\}$ and $p_\mu=\left\{p_0,p_1\right\}$, with the ordinary symplectic structure given by
\begin{align}
&\left\{x^\mu,x^\nu\right\}=0,\notag\\ 
&\left\{x^\mu,p_\nu\right\}=-\delta^\mu_\nu,\\
&\left\{x^\mu,x^\nu\right\}=0.\notag
\end{align}
the generators are represented, at first order in $\ell,H$ and $H\ell$, by
\begin{align}
&P_0=p_0-H x p,\notag \\
&P_1=p_1,\\
&N=p_1 t+p_0 x-H\left(p_1 t^2\frac{p_1 x^2}{2}\right)-\ell x\left(p_0^2+\frac{p_1^2}{2}\right)+H\ell p_1 x\left(p_1 t+\frac{3}{2}p_0 x\right),\notag
\end{align}
and the Casimir reads as
\begin{equation}\label{qcasimir}
C_{q\text{dS}}=p_0^2-p_1^2\left(1+\ell\,p_0\right)\left(1-2\,H\,t\right).
\end{equation}
From the expression above, as previously anticipated,  taking the limit $H\rightarrow 0$ one recovers the Casimir of the $\kappa$P algebra, while in the limit $\ell\rightarrow 0$ the Casimir of the classical de Sitter algebra is obtained.

\subsection{Finsler spacetime from the q-de Sitter mass Casimir}\label{fqdS}

We start by considering the action of a free particle with a constraint imposing the mass shell condition in terms of the Casimit \eqref{qcasimir} and it is given by
\begin{equation}\label{action}
I=\int \left[\dot{x}^\mu p_\mu-\lambda(\tau)\left(C_{q\text{dS}}-m^2\right)\right]d\tau,
\end{equation}
where $\lambda(\tau)$ is a lagrange multiplier enforcing the on-shell condition that we rewrite as
\begin{equation}
C_{q\text{dS}}=m^2\,\rightarrow\,p_0^2=m^2+a^{-2}(t)p_1^2\left(1+\ell\,p_0\right),
\end{equation}
where $a(t)=e^{Ht}=1+Ht+\mathcal{O}(H^2)$ is the classical dS scale factor.

The associated equations of motion are given by
\begin{subequations}\label{hameq}
\begin{align}
& \dot{t}=\lambda\,\left[2p_0-\ell\,a^{-2}p_1^2\right],\\
& \dot{x}=-2\lambda\,a^{-2}\,p_1\left(1+\ell p_0\right),
\end{align}
\end{subequations}
and they can be inverted to give\footnote{Assuming $\lambda\sim\mathcal{O}(1)$.}
\begin{subequations}\label{invhameq}
\begin{align}
& p_0=\frac{\dot{t}}{2\lambda}+\ell\,a^{2}\frac{\dot{x}^2}{8\lambda^2},\\
& p_1=-\frac{a^{2}\,\dot{x}}{2\lambda}\left(1-\ell\frac{\dot{t}}{2\lambda}\right).
\end{align}
\end{subequations}
Therefore the Lagrangian in \eqref{action} written in terms of velocities and the Lagrange multiplier reads as
\begin{equation}\label{polyakov}
L=\frac{\dot{t}^2-a^{2}\,\dot{x}^2}{4\lambda}+\ell\frac{a^{2}\,\dot{t}\dot{x}^2}{8\lambda^2}+\lambda m^2.
\end{equation}
In the limit $a(t)\rightarrow 1$ we recover the Lagrangian in \cite{Amelino-Camelia:2014rga}, as expected. The Lagrangian above can be minimized with respect to $\lambda$ to give
\begin{equation}\label{lagrmult}
\lambda=\frac{1}{2}\frac{\sqrt{\dot{t}^2-a^{2}\,\dot{x}^2}}{m}+\frac{\ell}{2}\frac{a^{2}\,\dot{t}\dot{x}^2}{\dot{t}^2-a^{2}\,\dot{x}^2}.
\end{equation}
The Lagrangian \eqref{polyakov} can now be written in terms of velocities only and it reads as
\begin{equation}\label{lagr}
L=m\left(\sqrt{\dot{t}^2-a^{2}\,\dot{x}^2}+\frac{\ell m}{2}\frac{a^{2}\,\dot{t}\dot{x}^2}{\dot{t}^2-a^{2}\,\dot{x}^2}\right).
\end{equation}
The expression above is of degree one in the velocities and therefore it defines the following Finsler norm\footnote{It is worth noticing that, while finishing this work, the paper \cite{Lobo:2016xzq} appeared. The authors arrive to a similar result working in conformal time instead of comoving time.}
\begin{equation}\label{norm}
F=\sqrt{\dot{t}^2-a^{2}\,\dot{x}^2}+\frac{\ell m}{2}\frac{a^{2}\,\dot{t}\dot{x}^2}{\dot{t}^2-a^{2}\,\dot{x}^2}.
\end{equation}

According to the relation \eqref{fmetrdef}, a Finsler metric can be derived from \eqref{norm} and it reads as
\begin{equation}\label{fmetric}
g^F_{\mu\nu}(x,\dot{x})=\left(
\begin{array}{cc}
 1+\frac{3 a^4 m \ell  \dot{t} \dot{x}^4}{2 \left(\dot{t}^2-a^2 \dot{x}^2\right)^{5/2}} 
 & \frac{m \ell a^4 \dot{x}^3\left(a^2 \dot{x}^2-4 \dot{t}^2 \right)}{2 \left(\dot{t}^2-a^2 \dot{x}^2\right)^{5/2}} \\
 \frac{m \ell a^4 \dot{x}^3\left(a^2 \dot{x}^2-4 \dot{t}^2 \right)}{2 \left(\dot{t}^2-a^2 \dot{x}^2\right)^{5/2}} 
 & -a^2+\frac{m \ell a^2\dot{t}^3 \left(2 \dot{t}^2+a^2 \dot{x}^2 \right)}{2 \left(\dot{t}^2-a^2 \dot{x}^2\right)^{5/2}} \\
\end{array}
\right).
\end{equation}
When $\ell\rightarrow 0$ the metric above reduces to the one of a classical de Sitter space in coordinate time and for $a(t)\rightarrow 1$ the Finsler metric assocaited with $\kappa$P is recovered. The norm \eqref{norm} and the metric \eqref{fmetric} satisfy all the identities of a proper Finsler spacetime introduced in Sec.\ref{finslersec}.

Using \eqref{lagrmult} one can rewrite \eqref{hameq} to get
\begin{subequations}\label{explinvhameq}
\begin{align}
& p_0=\frac{m \dot{t}}{\sqrt{\dot{t}^2-a^2 \dot{x}^2}}-\frac{\ell m^2 a^2 \dot{x}^2 \left(a^2 \dot{x}^2+\dot{t}^2\right)}{2 \left(\dot{t}^2-a^2 \dot{x}^2\right)^2},\label{genmom0}\\
& p_1=-\frac{m\,a^2\dot{x}}{\sqrt{\dot{t}^2-a^2 \dot{x}^2}}+\frac{\ell m^2 a^2 \dot{t}^3 \dot{x}}{\left(\dot{t}^2-a^2 \dot{x}^2\right)^2},\label{genmom1}
\end{align}
\end{subequations}
and the following relations can be found as well
\begin{subequations}\label{explhameq}
\begin{align}
&\frac{m \dot{t}}{\sqrt{\dot{t}^2-a^2\dot{x}^2}}=p_0+\frac{\ell a^{-2} p_1^2}{2 m^2}\left(a^{-2}p_1^2+p_0^2\right),\\
&\frac{m\,a \dot{x}}{\sqrt{\dot{t}^2-a^2\dot{x}^2}}=-a^{-1}p_1\left(1+\frac{\ell}{m^2}p_0^3\right).
\end{align}
\end{subequations}
Using the relations above one recovers the mass shell condition as
\begin{equation}
m^2=\left(\frac{m\,\dot{t}}{\sqrt{\dot{t}^2-a^2\dot{x}^2}}\right)^2-\left(\frac{m\,a\,\dot{x}}{\sqrt{\dot{t}^2-a^2\dot{x}^2}}\right)^2=p_0^2-a^{-2}\,p_1^2 (1+\ell p_0).
\end{equation} 
and the Finsler metric \eqref{fmetric} can be rewritten in terms of momenta as
\begin{equation}\label{fmetricp}
g^F_{\mu\nu}(x,p)=\left(
\begin{array}{cc}
 1+\frac{3}{2}\frac{\ell p_0 p_1^4}{m^4} 
 & -\frac{a}{2}\frac{\ell p_1^3\left(p_1^2-4p_0^2\right)}{m^4} \\
  -\frac{a}{2}\frac{\ell p_1^3\left(p_1^2-4p_0^2\right)}{m^4} 
 & -a^2+\frac{a^2}{2}\frac{\ell p_0^3\left(2p_0^2+p_1^2\right)}{m^4} \\
\end{array}
\right).
\end{equation}
By comparing \eqref{norm} with the Finsler norm associated with the $\kappa$P symmetries in \cite{Amelino-Camelia:2014rga}, it can be noted that the two are conformally related as in the classical case.

Using \eqref{explinvhameq} and \eqref{explhameq} it can be shown that the inverse metric satisfies the following relations
\begin{subequations}
\begin{align}
& g^{\mu\nu}_F(x,\dot{x})p_\mu (\dot{x})p_\nu (\dot{x})=m^2,\\
& g^{\mu\nu}_F(x,p)p_\mu p_\mu=p_0^2-a^{-2}p_1^2(1+\ell p_0).
\end{align}
\end{subequations} 

We have shown so far that a particle with the $q$dS mass Casimir can be described in terms of a Finsler geometry through the norm \eqref{norm} and the metric (\ref{fmetric},~\ref{fmetricp}) and we noticed that this structure is conformally related to the one of $\kappa$P introduced in \cite{Amelino-Camelia:2014rga}.\footnote{The analsysis of the Killing equation, needed to prove the full equivalence between the symmetries of the Finsler geometry compatible with the $q$dS mass Casimir and the the $q$dS Hopf algebra, is not among the objectives of this work. See however \cite{Lobo:2016xzq}.}

In the tangent space, the corrections to the ordinary Minkowski norm (or metric) are given by terms which are of the form $\ell m\,f(\dot{x})$ or $\ell m\,g(p/m)$ in momentum space, with $f$ and $g$ some functions of velocities and momenta respectively. These kinds of corrections are typical of \textit{rainbow gravity} scenarios \cite{Magueijo:2002xx} (see also \cite{Lobo:2016lxm}). Similar results where also found in \cite{Assaniousssi:2014ota,Torrome:2015cga} where the propagation of particle in a quantum geometry was analyzed and the deviations from the classical results were given in terms of a dimensionless \textit{non-classicality} parameter $\beta$, involving expectation values of the geometrical operators over a state of the quantum geometry, and functions of $p/m$, without an explicit dependence on any fundamental scale. In the framework presented here, the analogous parameter would be represented by the dimensionless combination $\ell m$, which makes manifest the presence of a fundamental scale.

In the following subsection, we will explicitly derive the worldline of a particle propagating on this Finsler geometry associated to the dispersion relation $C_{q\text{dS}}=m^2$ and we will study the associated Christoffel symbols.

\subsection{Christoffel symbols and geodesic equations}

Worldlines in Finsler geometry can be derived using Euler--Lagrange equations which is equivalent to computing the geodesic equations given by 
\begin{equation}\label{geodeq}
\ddot{x}^\mu+\Gamma^\mu_{\rho\sigma}(x,\dot{x})\dot{x}^\rho\dot{x}^\sigma=0,
\end{equation} 
once the parameter $\tau$ has be chosen to be affine. The Christoffel symbols are defined as in Riemannian geometry
\begin{equation}
\Gamma^\mu_{\rho\sigma}(x,\dot{x})=\frac{1}{2}g^{F\mu\nu}(x,\dot{x})\left(\partial_\rho g^F_{\sigma\nu}+\partial_\sigma g^F_{\rho\nu}-\partial_\nu g^F_{\rho\sigma}\right),
\end{equation}
but now they depend on the velocities through the metric tensor. 

They are explicitly given by
\begin{subequations}\label{conn}
\begin{align}
\Gamma^0_{00}=&\frac{3 H m \ell\dot{t} a^4\dot{x}^4\left(4 \dot{t}^2+a^2\dot{x}^2\right)}{4 \left(\dot{t}^2-a^2\dot{x}^2\right)^{7/2}},\\
\Gamma^0_{01}=&\frac{H m \ell a^4 \dot{x}^3  \left(4 \dot{t}^2-a^2\dot{x}^2\right)}{2 \left(\dot{t}^2-a^2\dot{x}^2\right)^{5/2}},\\
\Gamma^1_{00}=&-\frac{H m\ell a^2 \dot{x}^3  \left(16 \dot{t}^4-2 a^2\dot{t}^2 \dot{x}^2+a^4\dot{x}^4\right)}{2 \left(\dot{t}^2-a^2\dot{x}^2\right)^{7/2}},\\
\Gamma^0_{11}=&Ha^2-\frac{1}{4}\frac{H m \ell a^2 \dot{t}  \left(4 \dot{t}^6+10 \dot{t}^4 a^2 \dot{x}^2+7 \dot{t}^2 a^4 \dot{x}^4-6 a^6 \dot{x}^6\right)}{\left(\dot{t}^2-a^2\dot{x}^2\right)^{7/2}},\\
\Gamma^1_{01}=&H-\frac{3 H m\ell a^2 \dot{t}^3 \dot{x}^2 \left(4 \dot{t}^2+a^2\dot{x}^2\right)}{4 \left(\dot{t}^2-a^2\dot{x}^2\right)^{7/2}},\\
\Gamma^1_{11}=&-\frac{H m\ell a^4 \dot{x}^3\left(a^2\dot{x}^2-4 \dot{t}^2\right)}{2 \left(\dot{t}^2-a^2\dot{x}^2\right)^{5/2}}.
\end{align}
\end{subequations}
In the limit $\ell\rightarrow 0$ they reduce to the Christoffel symbols of a classical de Sitter space while for $H\rightarrow 0$ they vanish in agreement with the fact that in this limit the Finsler metric of $\kappa$P is recovered. We also notice that the correction terms to the \textit{classical} results are proportional to the combination $H\ell$. 

With the parametrization $F=1$ applied to the norm \eqref{norm}, the geodesic equations are specifically given by
\begin{subequations}\label{fingeod}
\begin{align}
&\ddot{t}+H\,a^2\dot{x}^2\left(1-2\ell m\dot{t}\right)=0,\\
&\ddot{x}+H\dot{x}\left(2\dot{t}+\ell m a^2\dot{x}^2\right)=0,
\end{align}
\end{subequations}
and their dependence on the mass $m$ signals a violation of the WEP.

In order to explore the consequences of these corrections one can expand \eqref{conn} up to second order in $H$ obtaining
\begin{subequations}\label{expconn}
\begin{align}
\Gamma^0_{00}\simeq\,&3 H\ell  m\dot{t}\left(\frac{4 \dot{t}^2 \dot{x}^4+\dot{x}^6}{4 \left(\dot{t}^2-\dot{x}^2\right)^{7/2}}+\frac{H t\left(16 \dot{t}^4 \dot{x}^4+18 \dot{t}^2 \dot{x}^6+\dot{x}^8\right)}{4 \left(\dot{t}^2-\dot{x}^2\right)^{9/2}}\right),\\
\Gamma^0_{11}\simeq\,&H+2 H^2 t-H\ell m \dot{t}\left(\frac{4 \dot{t}^6+10 \dot{t}^4 \dot{x}^2+7 \dot{t}^2 \dot{x}^4-6 \dot{x}^6}{4 \left(\dot{t}^2-\dot{x}^2\right)^{7/2}}+\right.\\\nonumber
&\left.+\frac{Ht\left(8\dot{t}^8+60\dot{t}^6 \dot{x}^2+72\dot{t}^4 \dot{x}^4-41\dot{t}^2 \dot{x}^6+6\dot{x}^8\right)}{4 \left(\dot{t}^2-\dot{x}^2\right)^{9/2}}\right),\\
\Gamma^1_{01}\simeq\,&H-3H\ell m\dot{t}^3\dot{x}^2 \left(\frac{4 \dot{t}^2+\dot{x}^2}{4 \left(\dot{t}^2-\dot{x}^2\right)^{7/2}}+\frac{H t\left(8\dot{t}^4 +24\dot{t}^2 \dot{x}^2+3\dot{x}^4\right)}{4 \left(\dot{t}^2-\dot{x}^2\right)^{9/2}}\right),
\end{align}
\end{subequations}
and similarly for the other components. One finds terms that are purely of order $H\ell$ and others that are of order $H\ell Ht$. If $t$ is at most $\mathcal{O}(H^{-1})$, the second kind of corrections is never bigger than the first one and this is true also for the higher order corrections since they are all multiplier by coefficients of the type $H\ell (Ht)^{n-1}$. 

Therefore, if one neglects correction terms which are proportional to $H\ell$, the Christoffel symbols become independent of $\dot{x}^\mu$ and this condition is preserved as long as $t$ is not larger than $H^{-1}$. In this limit the Finsler structure associated to $q$dS is approximately of the Berwald type and the Christoffel symbols are the same of a classical dS spacetime.

What happens at the metric tensor in this limit? Expanding \eqref{fmetric} up to first order in $H$ one gets
\begin{equation}
g_{\mu\nu}^F (x,\dot{x})\simeq\left(
\begin{array}{cc}
 1+\frac{3\ell m\dot{t} \dot{x}^4}{2 \left(\dot{t}^2-\dot{x}^2\right)^{5/2}}+\frac{3 Ht\ell m\dot{t}\dot{x}^4 \left(4 \dot{t}^2+\dot{x}^2\right) }{2 \left(\dot{t}^2-\dot{x}^2\right)^{7/2}} & -\frac{\ell m\dot{x}^3 \left(4 (4 H t+1) \dot{t}^4-(2 H t+5) \dot{x}^2 \dot{t}^2+(H t+1) \dot{x}^4\right)}{2 \left(\dot{t}^2-\dot{x}^2\right)^{7/2}} \\
 -\frac{\ell m\dot{x}^3 \left(4 (4 H t+1) \dot{t}^4-(2 H t+5) \dot{x}^2 \dot{t}^2+(H t+1) \dot{x}^4\right)}{2 \left(\dot{t}^2-\dot{x}^2\right)^{7/2}} & -1+2Ht+\frac{\ell m\dot{t}^3}{2} \left(\frac{  \left(2 \dot{t}^2+\dot{x}^2\right)}{\left(\dot{t}^2-\dot{x}^2\right)^{5/2}}+\frac{Ht\left(4 \dot{t}^4+10 \dot{x}^2 \dot{t}^2+\dot{x}^4\right)}{\left(\dot{t}^2-\dot{x}^2\right)^{7/2}}\right) \\
\end{array}
\right).
\end{equation}
In the metric above the constant $H$ always comes together with the coordinate time $t$ and this is also true for higher order terms that would come with coefficients of the type $(Ht)^n$. Therefore, while at the level of the Christoffel symbols these terms can be neglected as long as $t\lesssim H^{-1}$, this is not true for the metric tensor as one would get terms which are of the same order as the terms of $\mathcal{O}(\ell)$ i.e., $\ell (Ht)^n\sim \ell$ for $t\sim H^{-1}$. The metric is, therefore, still of Finslerian type form.

Having said that, at first order in $H$ and $\ell$ and ignoring terms proportional to $H\ell$ not enhanced by a factor of $t$, the geodesic equations \eqref{geodeq} are now the same that one would obtain from a classical dS spacetime\footnote{Note that analogous conclusions can be obtained in the framework presented in \cite{PhysRevD.92.084053} under similar hypothesis.}. They are given by
\begin{subequations}\label{berwgeod}
\begin{align}
&\ddot{t}+H \dot{x}^2=0,\\
&\ddot{x}+2 H  \dot{t}\dot{x}=0,
\end{align}
\end{subequations}
where any dependence on the mass has disappeared. Comparing \eqref{berwgeod} with \eqref{fingeod}, it is clear that in the former case the additional mass dependent term behaves like a force carrying the particle away from the classical geodesic motion.

On the other hand, the chronometric structure will still be velocity dependent and it will contain information on both the fundamental scale $\ell$ and the curvature scale $H$. For example, in the equations above, the derivatives are performed with respect to an affine parameter. In this respect, with the usual definition of proper time, from the metric \eqref{fmetricp} one obtains
\begin{equation}
\Delta\tau=\int_{t_1}^{t_2} \sqrt{g_{00}^F}\,dt=\int_{t_1}^{t_2} \left(1+\frac{3}{2}\frac{\ell p_0\,p_1^4}{m^4}\right)dt=\Delta t\left(1+\frac{3}{2}\frac{\ell p_0\,p_1^4}{m^4}\right)
\end{equation}
where we chose $dx=0$, so that no other components of the metric need to be considered and $p_0=const$ (in this frame there are no effects associated with $H$). Therefore the proper time turns out to be momentum (or velocity) dependent and particles with different energy will experience different elapsed proper time intervals $\Delta\tau$, given the same coordinate time interval $\Delta t$.

Let us now compute the trajectory of a particle as a function of coordinate time to show that indeed the non trivial structure of momentum space is not lost. Since the Lagrangian \eqref{lagr} does not depend on the spatial coordinate $x$, Euler--Lagrange equations tell us that the generalized momentum \eqref{genmom1} is conserved, i.e., $\dot{p_1}=0$. Therefore eq.~\eqref{genmom1} can be integrated, in the gauge $\tau(t)=t$ with the condition $x(0)=0$, and the result is given by
\begin{equation}\label{geodesic}
x(t)=\frac{p_1 t}{\sqrt{p_1^2+m^2}}\left[1-\frac{H t}{2}\left(1+\frac{m^2}{p_1^2+m^2}\right)\right]-\ell p_1 t\left(1-H t\right),
\end{equation}
for an incoming particle. The derivative of \eqref{geodesic} gives the speed of propagation that reads as\footnote{This result is in agreement with what has been found in \cite{Barcaroli:2015eqe,Lobo:2016xzq}.}
\begin{equation}
v(t)=\frac{p_1}{\sqrt{p_1^2+m^2}}\left[1- H t\left(1+\frac{m^2}{p_1^2+m^2}\right)\right]-\ell p_1(1-2 Ht)\quad\xrightarrow{m^2\rightarrow 0}\quad v(t)=1-H t-\ell p_1(1-2 Ht).
\end{equation}

Before going to the conclusion, let us briefly recap the results of this section. Eq.s \eqref{expconn} show that, in general, the $q$dS Finsler geometry is not of the Berwald type, since the spray coefficients (defined in Sec.~\ref{berwald}) are not quadratic in the velocities. However, it turns out that, in the specific regime $t\lesssim H^{-1}$, the Christoffel symbols become velocity-independent, and identical to the ones of a classical dS spacetime, and the Finsler geometry is approximately of the Berwald type.\footnote{Taking this limit is equivalent to ignore correction terms proportional to $(H\ell)^n$ which are not enhanced by a factor of $t^n$.}. Yet, the chronometric structure of the model does not become classical and the non trivial structure of the Finsler metric is maintained.

\section{Conclusions and outlook}

In this work we extended the relationship between theories with deformed relativistic symmetries and Finsler geometry by including the presence of spacetime curvature. In the first part, we have shown that the propagation of particles with deformed de Sitter symmetries, given by the $q$dS Hopf algebra, can be described in terms of a velocity and coordinate dependent of Finsler norm and we noted that the latter is conformally related to the $k$P Finsler norm introduced in \cite{Amelino-Camelia:2014rga}. Then, we studied the affine structure of the model by computing the generalized Christoffel symbols and pointing out that in general they remain velocity dependent. This allowed us to conclude that the $q$dS Finsler spacetime is not in general of the Berwald type and therefore the WEP is violated.

Nevertheless, we have shown that when the correction terms proportional to $H\ell$ (the product of the inverse of the curvature scale and the fundamental length scale) can be disregarded, the affine structure become classical, at least for a time scale which is at most comparable with the Hubble time $H^{-1}$. In this limit the Finsler structure becomes of the Berwald type and the WEP is recovered. On the other hand, in the same regime, the chronometric structure does not become completely classical. Indeed, the typical DSR effects, such as momentum dependent speeds of propagation for massive and massless particles, are still present and they come with both Planck scale and curvature corrections.

Deformations of the standard Poincar\'{e} algebra have been largely considered in the literature in the last twenty years but they are mostly used to described kinematical properties of particles with modified dispersion relations in a well defined relativistic framework. Whether these symmetry groups can be used to construct families of \textit{momentum dependent} (metric) theories of gravity, which would modify GR incorporating some QG features, is currently an open question. In the absence of concrete and realistic proposals for such kinds of theories, the study of deformed symmetry groups of non-flat spacetimes is a first step in understanding if such theories can be constructed.

As we anticipated in the introduction, two fundamental ingredients of any metric theory of gravity are LI and the WEP. The former can somehow be extended to include deformed symmetry groups and we have shown that indeed the $q$dS Finsler spacetime locally reduces to the flat $\kappa$P Finsler spacetime introduced in \cite{Amelino-Camelia:2014rga}. Therefore one may think of building a theory of gravity whose solutions locally look like a flat spacetime with $\kappa$P symmetries e.g., the $\kappa$P Finsler spacetime. However, the WEP is broken in $q$dS. Indeed, we found that the corrections to the ordinary geodesic equations come with a mass dependence. 
This additional component is negligible in the limit of small curvature and for typical time scales smaller than the Hubble time. In this limit the Finsler structure associated with $q$dS becomes of the Berwald type, which represents a subclass of Finsler spaces for which free falling (Fermi) normal coordinates can be defined and any free falling observer looking at neighboring free falling particles observes them moving uniformly over straight lines (formally implementing the idea of the \textit{Einstein's elevator}). Therefore, comparing the geodesic equations in this limit to the ones obtained without any approximation, we realized that the correction terms can be interpreted as force-like contributions. 

The most stringent bounds on violations of the WEP come from high precision E\"{o}tv\"{o}s-type experiments, but they are mostly performed in the gravitational field produced by the Earth and for macroscopic, composite bodies (see \cite{lrr-2006-3} and references therein). The relevant parameter used to constrain violations of the WEP is the so called \textit{E\"{o}tv\"{o}s ratio} $\eta$ that measures the fractional difference in acceleration between two bodies and it is currently bounded to be less or equal than about $10^{-13}$. Obviously, this bound cannot be directly applied to the present framework and tests of the WEP on cosmological scales would be more appropriate.

On the other hand, assuming that today's total energy density can be completely associated with the cosmological constant and that the universe is described by the $q$dS Finsler geometry, one can try to estimate how good is the Berwald approximation. Today's value of the Hubble parameter is approximately given by $H_0\simeq68\,$(km/s)/Mpc which corresponds, in seconds, to $H_0\simeq 2.2 \times 10^{-18}\,\mbox{s}^{-1}$. Assuming that $\ell$ is of the order of the Planck length $\ell_P\simeq 1.6\times 10^{-35}\,$m, the dimensionless combination $\ell H$, in natural units, is given by $\ell H\simeq 3.7\times 10^{-62}\ll 1$. Since this is the combination driving the correction terms in the geodesic equations, we expect the violation of the WEP to be very much suppressed in this context. 

At this point, one may wonder whether the effective gravitational dynamics for this theory can be described in terms of a sort of metric-affine theory of gravity\footnote{See e.g. \cite{Vitagliano:2010sr} for background material.} (at least for a time scale $t\lesssim t_H$), where the connections are the ones associated with a classical dS spacetime while the chronometric properties are given by the velocity dependent Finsler metric of $q$dS. In this case the Ricci tensor would be constructed solely on the basis of the classical connections and the Ricci scalar would be the contraction of the Finsler metric with the Ricci tensor. Still, it would be interesting to have a definite model providing such a dynamics.

Finally, one can also speculate that similar effects would be present in some kind of $\kappa$P-like deformation of the spherically symmetric gravitational field generated by a mass $M$. This would actually provide a framework to realistically test DSR models through tests of the WEP, as a bound on $\eta$ could imply a bound on the fundamental scale $\ell$.~\footnote{In the limit in which the Finsler structure is of the Berwald type, we do not consider energy dependent velocities as sources of WEP violations because particles with the same masses and same initial velocities will (approximately) experience the same acceleration.}
We hope to further develop these themes in future works.

\begin{acknowledgments}
The authors wish to acknowledge the John Templeton Foundation for the supporting grant \#51876. ML would like to thank Alessio Belenchia for interesting discussions during the preparation of this work.
\end{acknowledgments}

\bibliographystyle{ieeetr}
\bibliography{berwald_biblio}

\end{document}